# TOTAL FACTOR PRODUCTIVITY AND ITS DETERMINANTS: AN ANALYSIS OF THE RELATIONSHIP AT FIRM LEVEL THROUGH UNSUPERVISED LEARNING TECHNIQUES

**Paolo Pedotti (University of Milan & University of Pavia)**

**Abstract.** The paper is related to the identification of firm's features which serve as determinants for firm's total factor productivity through unsupervised learning techniques (principal component analysis, self organizing maps, clustering). This bottom-up approach can effectively manage the problem of the heterogeneity of the firms and provides new ways to look at firms' standard classifications. Using the large sample provided by the ORBIS database, the analysis covers the years before the outbreak of Covid-19 (2015-2019) and the immediate post-Covid period (year 2020). It has been shown that in both periods, the main determinants of productivity growth are related to profitability, credit/debts measures, cost and capital efficiency, and effort/efficiency of the R&D activity conducted by the firms. Finally, a linear relationship between determinants and productivity growth has been found.

## 1. INTRODUCTION

### *Motivation*

One of the big issues of economics is represented by the studying of the economic growth of countries, firms, and individuals, and by the understanding of the drivers which are related to.

There is a huge body of literature that contributes to the development of theories and models in this direction, starting from classical and neoclassical contributions. The crucial point from the first studies by these economists is related to the concept of productivity, in particular total factor productivity (TFP). More recent works, such as Cardona, Kretschmer and Strobel (2013), return on this point looking at productivity as the main driver of economic growth and on its impact the individuals (not casually they refer to productivity as the "wealth of nations"). But where does exactly productivity originate?

In this paper, I assume that the main source of productivity is given by the outcome of the research and development activity by the firms, measured in terms of intangible assets. This view is coherent with the Growth accounting literature (Solow, 1957; Romer, 1990), but also with many empirical studies. In this sense, the more efficient use of resources (a technological improvement) allows to obtain higher long-term economic growth.

The paper analyzes the relation between productivity and its determinants at the firm level. The aim consists of looking from a different perspective at this relation, since there is a huge heterogeneity in the data at the firm level that is only partially captured by neoclassical models and standard econometric techniques (Syverson 2004; Dosi, Grazzi, Tomasi and Zeli, 2012; Maue, Burke and Emerick, 2020).

***Research questions***

Overall, much work has been done on the link between TFP and some specific determinants (Melitz, 2003; Acemoglu, Aghion, & Zilibotti, 2006; Del Gatto, Ottaviano, & Pagnini, 2007; Melitz & Ottaviano, 2008; Aw, Roberts, & Yi Xu, 2008; Doraszelski & Jaumandreu, 2009; Faggio, Salvanes & Van Reenen, 2010; Syverson, 2011; Cardona, Kretschmer, & Strobel, 2013; Berthou et al, 2016; Bottazzi, Grazzi, Secchi & Tamagni, 2017; Pieri, F., Vecchi, M. & Venturini, F., 2018; De Loecker, Eeckhout et al., 2018 & 2020; Bajgar et al., 2019; Philippon, 2019; McAdam et al., 2019; Covarrubias et al., 2020; Gutierrez & Philippon, 2023).

Here, I provide a complementary approach in which I look at the association between TFP growth and a large set of accounting variables using unsupervised learning approaches (PCA, SOM and k-means clustering). These techniques allow indeed to consider firm heterogeneity working directly on firms' features as emerged by data analysis in two periods: pre- and post-Covid.

Indeed, while most of the previous works in this field focus on understanding how one specific determinant impacts firms' productivity, this study provides new ways to look at the relationship between TFP growth and all the determinants. Thanks to this bottom-up approach, the paper tries to explain how firms' heterogeneity and thus firms' characteristics impact both firms' classification and the strength of the relation between TFP and determinants. Finally, the introduction of a principal component regression provides a way to evaluate potential patterns of TFP growth.

***Content of this paper***

The analysis follows several steps. At first, the paper considers two main issues widely discussed in the literature: measurement issues and applications. Attention has been given to the selection of the best possible methodology for estimating the total factor productivity.

Then, some accounting variables are considered for the aim of representing firms' characteristics. Some of them are directly referring to some determinants: capital efficiency (i.e. property, plant & equipment per worker), firms' profitability (i.e. net profit per worker), R&D effort/efficiency (i.e. R&D expenditures and intangibles per worker), credit/debt measures (i.e. short-term debt per worker).

The result of the principal component analysis allowed the identification of at least 3 common principal components (PCs) for the two periods which can be interpreted as capital efficiency, profitability and cost management, and effort/efficiency in R&D activity.

After having discussed the main determinants through Self-Organizing Maps (SOM), the output of the SOM has been used as input for performing a clustering analysis to group firms according to PCs and TFP growth estimates: several groups have been found in this way in both periods and are mainly based on the average productivity growth. However, there is at least one group in both the pre- and post-Covid periods that shares the same average TFP growth, denoting a potential role for specific determinants.
Finally, the principal component regression results identify a linear relationship between PCs and productivity growth. This relation is checked also when sectorial and geographic controls are introduced. However, it does not hold if the principal component regression is performed within each cluster.
The rest of the paper is organized as follows: section 2 is related to the literature review, both on measurement issues (2.1) and on applications (2.2); section 3 introduces the data description and the methodologies of the analysis (in particular, section 3.1 is related to data presentation; section 3.2 to the estimation of TFP growth; section 3.3 to PCA methodology; section 3.4 to self-organizing maps and cluster analysis methodology and section 3.5 to principal component regression and Lasso); section 4 discuss the empirical analysis both on TFP growth estimation (4.1), principal component analysis (4.2), SOM and clustering (4.3) and principal component regression and Lasso (4.4); finally, section 5 resumes the main results and concludes the paper with some policy recommendations.

## 2. PRODUCTIVITY: A LITERATURE REVIEW

For the aim of this paper, I refer to productivity using the concept of Total Factor Productivity (TFP).[1] TFP is mainly described in the literature as the ratio between economic output and the combination of inputs required for production. Intuitively this result is associated with the level of technology adopted by the unit of observation (firm, sector, or country). Indeed, as stated by Comin (2006):

> Total Factor Productivity (TFP) is the portion of output not explained by the amount of inputs used in production. As such, its level is determined by how efficiently and intensely the inputs are utilized in production.

An alternative formulation of this concept is the one provided by Altomonte & Di Mauro (2022) which identifies total factor productivity (or "multi-factor productivity") as the effectiveness of the production process to bundle together several inputs (intermediates, energy, labour, capital, etc…) to produce an output. While the first definition by Comin (2006) focuses more on the existence of something that exceeds the strict combination of inputs as the output of the production process, Altomonte & Di Mauro (2022) rely more on the effectiveness of the process itself.

---

[1] The notion of TFP differs from the one of Partial Factor Productivity which refers to a single factor of production (i.e. capital or labor). PFP may be general, as the output per unit of labor or capital, or specific, as the ratio between the economic output obtained by one single factor and the amount of that factor which is required for the production (Frigero, 1979).

Although these considerations, the two definitions are practically equivalent. Both are saying that the production process transforms inputs into outputs in such a way that a "value added" is produced. However, the way in which this "value added" is measured differs in many authors works according to different approaches, methodologies, and fields of research (Del Gatto, Di Liberto, & Petraglia, 2009; Cardona, Kretschmer, & Strobel, 2013; Altomonte & Di Mauro, 2022).

## 2.1 Productivity: a literature review on measurement issues

The main line of distinction in measuring productivity is typically represented by the implementation of different groups of methodologies.[2]

A first group of methods aims to obtain a deterministic measure of productivity through non-parametric techniques. The most popular and consolidated deterministic methodology is provided by the neoclassical approach on production theory and growth accounting techniques, whose Robert Solow's paper on technical change and production function represents a first main contribution (Solow, 1957). In this framework, TFP coincides with the Solow residual, and can be interpreted as the "residual measure of our ignorance" (Hulten, 2001).

A second important group of methodologies allows, instead, the estimation of productivity (in levels or growth rates) by using parametric or semi-parametric approaches, and econometric techniques.

In these cases, the regression is obtained through the log-linearization of the production function where the coefficients are the output elasticities relative to the inputs and several control variables may be also included. Even in this case, productivity is measured as Solow residual.

For studies conducted at the firm (or industry) level, it is important to introduce fixed effects since productivity may vary consistently across sectors due to unobserved heterogeneity in the data, as many factors correlated with productivity may have not been considered. In those cases, findings are related to intra-sectoral variation in productivity, allowing to compare firms that produce similar goods (Maue, Burke, & Emerick, 2020).

In addition, econometric techniques have the advantage to estimate the elasticities of output to input growth directly through the model. Thus, many neoclassical assumptions typical of the deterministic setting can be relaxed (Cardona, Kretschmer, & Strobel, 2013).

On the other side, the main disadvantage is represented by the endogeneity issue that arises when all the input factors are chosen by the firms (Draca, Sadun, & Van Reenen, 2006). Moreover, following these approaches implies that any consideration on the causality of the relationships is avoided since the results are just establishing correlations between variables.

[2] Considering that the list of methodologies may be quite large since there are also several ways to combine them in different research settings, here I am just providing the most used methods according to the well-established literature. Then, in the next paragraphs, many other settings may be discussed whenever they're helpful in explaining some determinants of productivity.

## 2.2 Productivity: a literature review on TFP studies

With respect to the applications, as the empirical part of this paper will discuss productivity looking at microeconomic data, we are mostly interested in discussing microeconomic studies which deal with issues regarding the total factor productivity at an individual level (firm or plant).[3]

### *Main methodologies*

A lot of studies introduced semi-parametric methods to estimate productivity at a firm level. These methodologies are based on the use of proxy variables as dependant variables, once the TFP relation is inverted. This is the case of intermediate goods as a function of TFP and capital as reported by Levinsohn and Petrin (2003) or the use of investments as proposed in Olley and Pakes (1996) to deal with the "invertibility condition" of the TFP (Del Gatto, Di Liberto & Petraglia, 2009).

These methods allow to find evidence on firms' productivity distribution and dispersion, which comes mainly from the heterogeneity of firms, but also from specific sectoral and industrial factors (Dosi, Grazzi, Tomasi and Zeli, 2012; Maue, Burke and Emerick, 2020), and even on which economic conditions may generate differences in productivity across firms. On this point, Syverson (2011) provides a classification of TFP determinants, identifying two categories: external and internal drivers. External drivers are those that come from the operating environment which is not directly controlled by the firms, such as productivity spillovers, trade, market and intra-market competition.

Internal drivers, instead, are related to all the factors that are directly controlled by the firms, such as the presence of managerial practice/talent, higher quality in labour and capital inputs, IT and R&D activity. For the purposes of this paper, the following sections provide an overview of the literature on the internal drivers of productivity as classified by Syverson's (2011), while external drivers are not considered. Although the empirical analysis does not capture the role of all the internal drivers that are presented here, the objective is to offer a sufficiently comprehensive review to make the reader understand the methodological advantages and limitations of the approaches employed.

### *Labour misallocation*

According to Altomonte & Di Mauro (2022), labour misallocation is characterised by three main business dynamics: the connection between job and worker reallocation; the heterogeneity of employment dynamics at the firm level; the cyclical behaviour of the labour market.

In particular, the continuous process of job creation and destruction, linked with firm exit/entrance in the market, was first discussed by Bartelsman and Doms (2000), based on Davis and Haltiwanger's

[3] For completeness, macroeconomic studies emphasize the role of TFP on the dynamics of the economics growth at an aggregate level, with particular attention to differences that occur between countries (Del Gatto, Di Liberto, & Petraglia, 2009).

(1992) work on job flows across plants, and by Bartelsman et al. (2004) across several countries. They found this process to affect productivity growth, reallocating resources from the less productive firms to the most productive ones. Moreover, other factors, like business dynamism, market regulation and restrictions on foreign direct investments play a particular role in avoiding barriers to firm growth and resource allocation (Decker et al., 2014; Andrews and Cingano, 2014; Andrews et al., 2016).

***Capital misallocation***

Moving the attention to capital misallocation, one of the most important problems is represented by the presence of financial frictions. These frictions prevent firms from accessing external financial resources when they are requiring to, due to not enough internal funds.

Duval et al. (2017) analyse how firms' financial fragility and credit tightness at the country level may explain the slowdown of productivity growth in recent years (more financially constrained firms experienced the harder impact). Even Altomonte et al. (2018) find that a positive exogenous liquidity shock increases the overall amount of intangibles according to the data of French firms in the years of the financial crises. Financial constraints, thus, play an important role in this situation. These results are also coherent with a successive study by Besley et al. (2020), that introduces a measure for credit frictions based on firms' default probability.

***Role of innovation***

Looking at firm's innovation, there are several studies based on the idea that firms introduce innovation mainly through the activity of R&D. Moreover, the contribution of Crepon, Duguet and Mairesse (1998) provides evidence that firm's R&D activity can be directly associated with productivity growth.

Using micro data on French manufacturing firms they were able to derive a structural model that links productivity through innovation output and innovation output through R&D activity. In other words, firms invest in R&D expenditures as input for their innovation activity (two choices: investment decision, size), then the innovation process transforms inputs into outputs (measured as patents applications, matched to firm data) and, finally, innovation outputs contribute to firm's productivity.

The results of the analysis show that only 11% of the firms in the sample are R&D investing firms, and almost the same percentage (12%) are patenting firms with the half of them that has two or less patents. Moreover, innovation firms seem to be larger, more productive, and more capital intensive. These results are in line with Cohen and Klepper's (1996) analysis of the relation between firm size and R&D innovation process by the firms.

Larger firms have a higher probability to conduct R&D activities, while at the same time, their R&D intensity is not directly affected by. On the contrary, market shares, diversification, demand pull, and technology push indicators impact directly firms' R&D efforts. With respect to the patenting firms, firm size does not impact innovation output which is affected only by the research effort, demand pull and technology push indicators. As a final result, productivity is affected positively by higher innovation output (Crepon, Duguet and Mairesse, 1998).

***ICT capital, R&D expenditures and the organizational framework***

Following different approaches, some authors have studied the role of firm's innovation activity, analysing information and communication technology as a new type of capital, and evaluating its impact on productivity (Cardona, Kretschmer, & Strobel, 2013; Faggio, Salvanes & Van Reenen, 2010). On the other side, other authors are instead more sceptical of the role of IT in reshaping the economy.[4]

At the firm level, however, the activity of R&D explains a good amount of productivity growth. As it happens with IT capital, this is also coupled with an increase in the uncertainty of its outcome (Doraszelski & Jaumandreu, 2009). The positive effect on firm's productivity by the R&D expenditures is confirmed also by Pieri, Vecchi & Venturini (2018).

Looking at the organizational framework and profitability, innovation has been proved to be favoured by the selection of high-skills managers and firms (Acemoglu, Aghion, & Zilibotti, 2006), while productivity and profitability seem to be not so strongly related (Bottazzi, Grazzi, Secchi and Tamagni, 2017). Finally, in more recent years, the discussion has moved to the impact of Artificial Intelligence (A.I.) on economic activities (Aghion, Jones, & Jones, 2018; Brynjolfsson, Rock and Syverson, 2021).

## 3. METHODOLOGY AND DATA DESCRIPTION

This section is related to the presentation of the dataset used for the empirical analysis and to the description of the implemented methodologies.

After having provided a discussion on the accounting variables that are selected for the next steps, a framework that connects the deterministic models to the semiparametric techniques is introduced for the productivity growth estimation. Further on, unsupervised machine learning techniques are implemented: principal component analysis (PCA) is required to reduce the number of accounting variables that are considered, while self-organizing maps (SOM) and clustering analysis are used for developing new ways to look at firms' heterogeneity accordingly to firms' micro characteristics and their TFP growth.

Finally, PCs and TFP growth estimates are linked using Principal Component Regression (PCR) to provide further economic interpretation of the results in terms of TFP determinants (a comparison with the Lasso is included). The different steps of the analysis and their order are shown in Figure 1.

[4] Carr (2003) have looked to IT as a commodity (with the economic implication that is more efficient to "spend less" on it), and Gordon (2010) have discussed the presence of a diminishing return's problem.

*Figure 1 - Steps of the empirical analysis.*

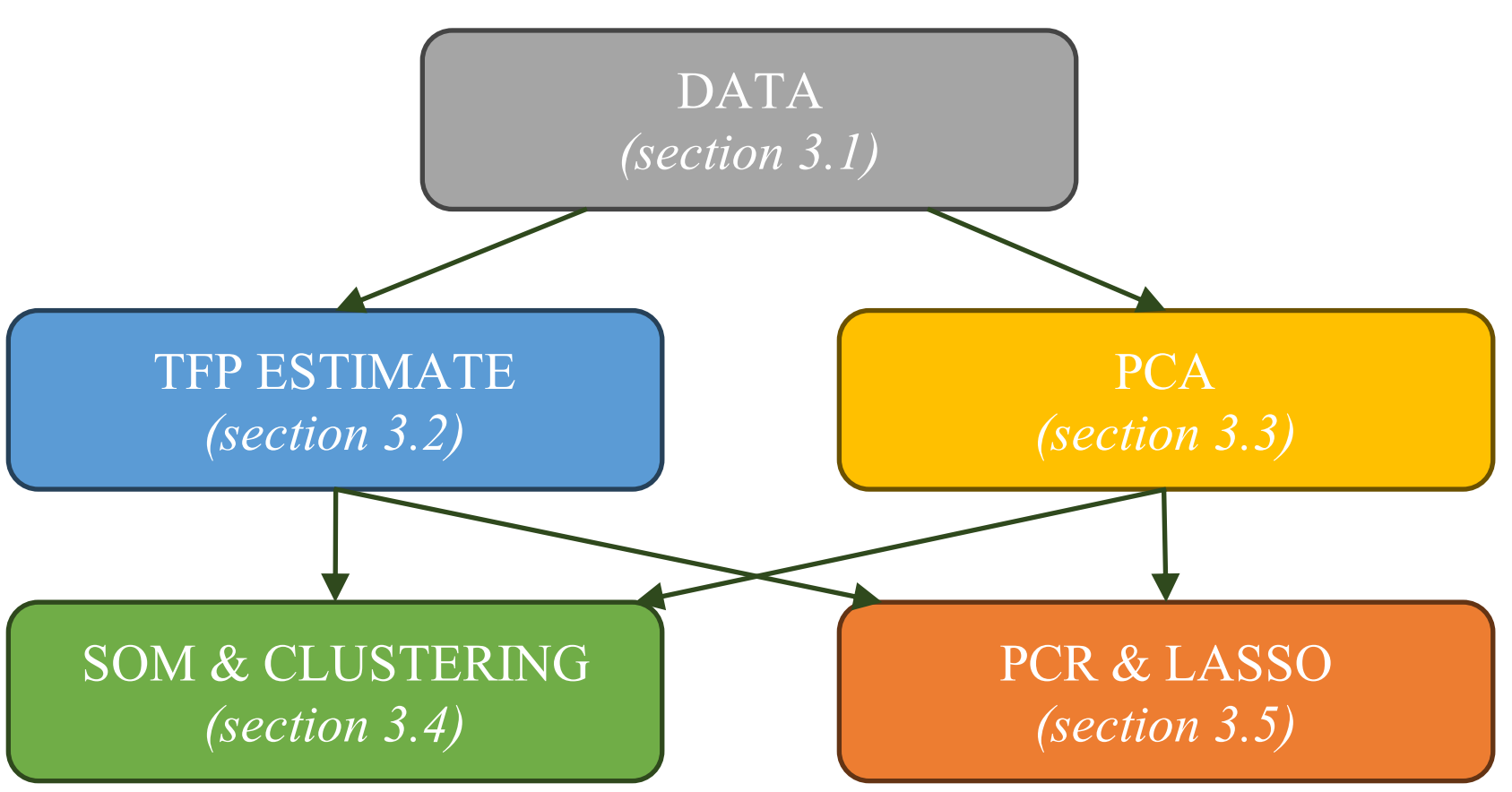

### 3.1 Data Description

The analysis is conducted at the firm level and considers the large sample provided by the database of ORBIS, which includes the data for the balance sheets of over 41 million firms. In particular, the final sample is obtained considering only corporate firms with any ownership, from any country and sector with the availability of data for gross sales, R&D expenses, and number of employees for at least the years 2019 and 2020 (19.852 firms collected).

Looking at some descriptive statistics, the economic variables used for obtaining productivity growth estimates are described in Tables 1 and 2 for the years before the outbreak of Covid-19 (the averages between 2015 and 2019) and for the post-Covid period (year 2020).

We may have an idea of the size of the firms in the sample by looking at the "workers" variable, which denotes the number of employers hired by each firm.[5] It emerges that our sample is quite heterogeneous in size: even large firms (more than 250 workers) are distributed over a large interval. Moreover, the sample average is slightly greater than the 85th percentile, which means that the sample average is strongly influenced by values from very large firms.[6]

All the other variables are represented by final goods (the output of the production by firms), intermediate goods, but also fixed assets, as a measure of the stock of capital owned by the firms, and short-term and long-term investments. All these variables show mean values greater than the 85$^{th}$ percentile of their distributions which means that larger firms have huge amounts of resources with respect to the "average" firm.[7]

---

[5] According to the EU standard classification for small and medium-sized enterprises (SMEs) micro firms have fewer than 10 workers, small firms have from 10 up to 50 workers, while medium firms have from 50 up to 250 workers. SMEs represent 99% of all the business in the US and in the EU (EU recommendation 2003/361).

[6] Note that the average number of workers per firm has increased (+182) from the pre-Covid period to the post-Covid period and even the 85% of the firms now have more workers on average (+114).

[7] Figure A.1 in the appendix shows densities and boxplots for some of the main variables (workers, final goods, fixed assets, investments) in both the pre- and the post-Covid period. Note that there are no significant changes in densities.

***Table 1 – Descriptive statistics for main variables (USD), averages 2015-2019.***

| **Variable** | **N** | **Mean** | **Pctl. 85** | **Max** | **Std. Dev.** |
|---|---|---|---|---|---|
| *final goods* | 14302 | 98282.55 | 72698.49 | 34739997.41 | 620897.91 |
| *workers* | 19852 | 4319.31 | 4107.35 | 670682.50 | 21029.44 |
| *fixed assets* | 19852 | 1380157.85 | 768210.47 | 305091875 | 9085822.87 |
| *short-term investments* | 8878 | 145921.15 | 70226.88 | 113345800 | 1865081.49 |
| *investments* | 11530 | 115261.65 | 44944.38 | 161069800 | 1981840.01 |
| *intermediate goods* | 10482 | 55406.11 | 31675.89 | 66119091.52 | 750283.98 |

***Table 2 – Descriptive statistics for main variables (USD), year 2020.***

| **Variable** | **N** | **Mean** | **Pctl. 85** | **Max** | **Std. Dev.** |
|---|---|---|---|---|---|
| *final goods* | 12959 | 111431.87 | 84308.02 | 33382009.34 | 697831.20 |
| *workers* | 19852 | 4501.41 | 4221.35 | 1298000 | 22955.57 |
| *fixed assets* | 19852 | 1675396.88 | 962330.21 | 370792595.31 | 10770286.26 |
| *short-term investments* | 5815 | 241349.25 | 108701.77 | 122951000 | 2683990.49 |
| *investments* | 7832 | 133480.78 | 56017.74 | 126809189.53 | 2011527.72 |
| *intermediate goods* | 9434 | 70103.97 | 37858.59 | 126749624.68 | 1361964.96 |

We can now move to the discussion of the accounting variables included in the dataset, which are then used for the PCA (descriptive statistics are presented in Tables A.1 and A.2 in the appendix).

A first group of variables is related to the assets section of the balance sheet which includes current assets (total value, inventories, account receivables, cash and other current assets) and long-term assets (net property, plant and equipment, goodwill and intangibles assets and other fixed and long-term assets). Another group of variables consider current liabilities as their total value, account and notes payables to creditors and other short-term debt, and long-term liabilities (interests and stock of total long-term debt). The last group is then related to the stockholder's equity, an accounting measure of the firm's net worth (Berk & DeMarzo, 2017), and to the total enterprise value.

All these variables express the financial position of the firms in terms of investments, debts and cash reserves and thus they are impacted over time by firm's decisions. Comparing mean values with the 85th percentiles of the distributions indicates that firm size is relevant for all variables. For this reason, the analysis considers variables in per-worker terms.

In the appendix (Tables A.3 and A.4) are reported even the accounting variables related to the income statement, which considers firm's revenues (total revenues, gross/net sales and all the measures of earnings and incomes), and expenses (cost of sales, R&D expenditures, general and administrative expenses, depreciation and amortization, financial expenses, tax and dividends) over time.

### 3.2 First step: TFP estimation

### 3.2.1 Model Description

The starting point for deriving a useful relation for the TFP is to consider the neoclassical production function where output is a function of observable inputs (labor, physical capital, and intermediate goods) and a Hicks neutral technology.[8]

At time t the production function for any firm i can assume a Cobb-Douglas specification form as

$$Y_{it} = A_{it}\, L_{it}^{\alpha l}\, K_{it}^{\alpha k}\, M_{it}^{\alpha m} \tag{1}$$

where $A_{it}$ is the level of technology, $L_{it}$ is the labor input (number of workers), $K_{it}$ is the physical capital and $M_{it}$ is the stock of intermediate goods.

Moreover, from (1) is also possible to derive an expression[9] equivalent to the econometric specification:

$$output_{it} = \alpha_1\, labor_{it} + \alpha_2\, capital_{it} + \alpha_3\, interm_goods_{it} + \varepsilon_{it} \tag{2}$$

where the residuals catch the productivity growth as Solow residuals.

Although this specification may be quite intuitive for representing our setting, it doesn't consider several issues, such as the selection and simultaneity problems.[10] Indeed, estimating the production function with standard econometric techniques (O.L.S., panel regressions) when firms' choices on serving the market and on input quantities are based on their productivity may generate biases (Olley and Pakes, 1996). As we want to obtain consistent estimates, these methods are then discussed and implemented.

### 3.2.2 Semi-parametric methods

The first method to be presented is the one provided by Olley and Pakes (1996, OP from now on).

The main assumption of the OP model is based on the existence of one unobserved state variable – firm's productivity – which describes firm's behaviour over time allowing for idiosyncratic changes. Moreover, profits are determined by a function of firms' state variables (age, physical capital and efficiency index, $\omega_{it}$) and factor prices.

At the beginning of every period, each firm decides to remain or leave the market. If it remains then it chooses labor and investments, which are used to determine the level of capital in the the next period as

$$k_{t+1} = (1-\delta)k_t + i_t\,. \tag{3}$$

---

[8] The derivation is based on Solow (1957) and slightly different from Draca, Sadun and Van Reenen (2006) as it includes only intermediates as additional input.

[9] The derivation is made considering the production function in natural logarithm $ln(Y_{it}) = \alpha_0 + \alpha_l \ln(L_{it}) + + \alpha_k\, ln(K_{it}) + \alpha_m\, ln(M_{it})$ and taking the first difference $\Delta ln(Y_{it}) = \Delta\alpha_0 + \alpha_l\, \Delta ln(L_{it}) + \alpha_k\, \Delta ln(K_{it}) + \alpha_m\, \Delta ln(M_{it})$ where $\Delta\alpha_0$ is the TFP growth and the other terms are in growth rates.

[10] Firstly described by Marschak and Andrews (1944).

Most importantly OP model assumes that the firm's efficiency index $\omega_t$ is known by the firm, and it evolves over time following an exogenous first-order Markov process. Since firms are supposed to maximize the expected discounted value of their future net cash flow, then investment and shutdown decisions should be generated by the perceived distribution of future market structures. Investment is described by the function

$$i_t = i_t(\omega_t, a_t, k_t) \qquad (4)$$

and is increasing in $\omega_t$ (at least when $i_t < 0$, Pakes, 1994).
This allows to obtain

$$\omega_t = h_t(i_t, a_t, k_t) \,. \qquad (5)$$

Let us consider the OP basic econometric specification (in logs) as

$$output_{it} = \beta_0 + \beta_a \, age_{it} + \beta_k \, capital_{it} + \beta_l \, labor_{it} + \omega_{it} + \eta_{it} \qquad (6)$$

where $\omega_{it}$ is unobserved productivity and $\eta_{it}$ is the unobserved and unpredictable shock to productivity. Now, the OP estimation algorithm assumes labor as the only variable factor (free variable), while capital and age are fixed (state variables) and affected by the distribution of $\omega_t$ conditional on the information at the previous period and on the values of $\omega_{t-1}$.
Thanks to the invertibility of the investment function (which serves as a proxy variable[11]), the unobserved productivity variable is indeed a function of observables and the specification may be rewritten as a semiparametric regression model (in logs):

$$output_{it} = \beta_l \, labor_{it} + \phi_t(i_t, a_t, k_t) + \eta_{it} \qquad \text{where} \qquad (7)$$

$$\phi_t(i_t, a_t, k_t) = \beta_0 + \beta_a \, age_{it} + \beta_k \, capital_{it} + h_t(i_t, a_t, k_t) \,. \qquad (8)$$

Then, the implementation of the algorithm allows to estimate $\beta_l$, $\beta_a$ and $\beta_k$ in two steps.[12]
The second method that is considered is the one by Levinsohn and Petrin (2003, LP from now on). In particular, they try to deal with the same issues – simultaneity and selection biases in the production function estimation – but using intermediate inputs as a proxy variable.[13]
Even the LP algorithm is characterized by the two-step approach used by OP. However, in this case, the first step determines the coefficients of $\beta_l$ and $\phi_t$, while the second step is required to compute both $\beta_m$ and $\beta_k$ (while in OP investments are not part of the production function).

---

[11] The use of proxy variables is aimed to control for the error component correlated with inputs (Levinsohn and Petrin, 2003).
[12] At first, the implementation of the partial linear model allows us to obtain the estimates of the freely variable inputs, $\beta_l$ and $\phi_t$; then, these estimates and the survival probabilities are used to compute $\beta_a$ and $\beta_k$. Please note that residuals $\eta_{it}$ are computed in the first stage and they're not a state variable (as $\omega_{it}$), thus they are not affecting firms' decisions.
[13] Intermediates are a valid proxy when several conditions are satisfied: the monotonicity condition must be satisfied, which means intermediates must be increasing in $\omega_t$ conditional on capital; markets must be perfectly competitive; the production technology used for intermediate inputs must be separable.

### 3.2.3 Further developments

Ackerberg, Caves, and Frazer (2015, ACF from now on) underline several identification problems that arise when OP and LP methods are used. In particular, with respect to the first stage of the algorithms, the labour coefficient can be estimated consistently if and only if the variability of the free variables is independent of the proxy variable. Otherwise, the coefficients estimated in the first stage would be perfectly collinear (Rovigatti and Mollisi, 2018). The ACF method introduces corrections for the functional dependence problem.

OP and LP methods can correctly identify $\beta_l$ just under very specific DGPs (their estimates may be quite precise although not consistent), while a consistent estimate can be obtained through the ACF method with less restrictive assumptions. For these reasons, I consider ACF as the main method for conducting the analysis, using OP and LP methods as further checks on the results.

## 3.3 Second step: unsupervised learning approach - PCA

The second step of the empirical analysis is related to the implementation of the principal component analysis (PCA). In particular, PCA allows data visualization in low dimensions with the highest possible variation given all the observations included in the dataset on the original regressors. This allows to identify new variables which can represent better firms' heterogeneity.

PCA, in fact, finds the first principal component (PC1) as the normalized linear combination of the features with the largest variance. Then, the second principal component (PC2) is found by the normalized linear combination with the highest variance among all the linear combinations uncorrelated with PC1, and so on and so forth for the other principal components.[14]

## 3.4 Third step: unsupervised learning approach - SOM & clustering

Self-organizing maps (SOM) are then introduced to identify potential clusters and to discuss their interpretation. They consist of several maps that allow to synthesize and better visualize a dataset with multiple dimensions in a two dimensional plane.

Afterward, the output of the SOM is collected for the clustering analysis. In particular, the K-Means algorithm is implemented to split the observation sample into several subgroups to obtain a new classification, based on firms' heterogeneity. Indeed, elements within each group are expected to be the most like each other, while they are expected to be the most different among groups.[15]

---

[14] Scree plots help in deciding how many principal components are computed according to the proportion of the variance explained.

[15] The number of clusters is pre-defined: the problem consists of minimizing the within-cluster variation W(Ck) where the functional of W is chosen to be the Euclidean distance. In practice, each observation is randomly assigned to a cluster, then centroids are computed, and observations are redistributed to the nearest centroid, centroids are re-computed, and so on.

### 3.5 Fourth step: PC and Lasso regression

The last step of the analysis corresponds to a further look at the relation between firm's total factor productivity, as estimated according to the semi-parametric techniques, and the principal components. In particular, the principal component regression can be used to identify the relationship between firms' productivity and the related predictors (in other words, its determinants). A robustness check is performed using the original variables as regressors in a shrinkage method, Lasso regression.[16]

## 4. EMPIRICAL ANALYSIS

This section relates to the implementation of the methodologies discussed in the previous paragraphs. The structure reflects the one of Section 3: the stages of the analysis are presented in depth and then the results are discussed. In the illustration of the results, particular attention is given to the economic meanings and interpretations of the different stages.

### 4.1 First step: TFP estimation

In the first step of the analysis, the production coefficient estimates are estimated according to the three semi-parametric methods (OP, LP and ACF). In addition, it is important to consider that the sample using the OP methodology is smaller than the one used for the other two methods.

Looking at the results, estimates using OP and ACF for $\beta_l$ (0.52 and 0.57) and estimates using OP and LP of the first stage coefficient $\phi_t$ (0.51 and 0.56) are not so different: it seems that the use of different samples has just affected the estimation of $\beta_l$ for LP (0.33). Slightly different estimates are obtained for the second stage coefficient $\beta_k$ (OP 0.34, LP 0.38, ACF 0.44).

Since the coefficient for $\beta_l$ is quite like the one obtained with OP, and $\beta_k$ estimate is only slightly larger we can confirm the use of ACF as the beanchmark estimation method for this analysis. Moreover, differences in partial productivity estimates across methods should not influence the relation between TFP estimates and the other economic variables. The residuals collected at the end of the first stage are, indeed, strongly positively related and may serve as alternative TFP growth measures.[17]

### 4.2 Principal Component Analysis

The economic intuition underlying the use of the principal component analysis lies in finding new variables that can describe the characteristics of heterogeneous firms better than the original features.

---

[16] As other shrinkage methods, like the ridge regression, Lasso shrinks the coefficient estimates towards zero. In addition, a penalty term is introduced for forcing some estimates to be exactly zero and thus perfoming variable selection and enhancing the interpretability of the results (James, Witten, Hastie & Tibshirani, 2017).

[17] The results of the analysis based on OP and LP methods are available upon request.

Both the balance sheet and income statement variables, provided by the ORBIS dataset (see paragraph 3.1), are useful indicators of the way in which firms manage their productivity given financial and input constraints. Thus, PCA considers all those variables except the variables used for the TFP estimation.[18]
Looking at the results, reported in Tables 3 and 4, in both periods only the first three principal components explain more than 9% of the variance of the original variables (with some differences between the two periods). Moreover, these principal components are the only ones that capture almost the same positive and negative correlations with the original features in both periods.[19]
The first principal component is strongly associated with property, plant and equipment (net and other measures, positively), depreciation and amortization (negatively) and it may be seen as a proxy for the efficiency in the use of fixed assets in the production process. It may be interpreted as a measure of the optimal allocation of capital resources in the long run by firms (capital efficiency).
The second principal component is positively related again to property, plant and equipment measures, denoting a particular role for capital efficiency. But it is associated negatively with net profits and earnings after tax in both periods. Therefore, it may be seen as a measure of firm's profitability (inverse relation) and secondarily as efficiency in cost management (both related to fixed assets, and goods sold).
Finally, the third principal component is positively related to different credit/debt measures in the two periods (other short-term debt in the pre-Covid period and accounts receivable in the post-Covid period). Moreover, it's negatively associated with R&D expenditures and on property, plant and equipment. It measures different credit/debit measures, but it can be used as a proxy for firm's R&D effort/efficiency.

***Table 3 – correlations between principal components and features, pre-Covid period.***

| **(pre-Covid)** | **Var. Explained** | **Top Positive Correlated Variables** | **Top Negative Correlated Variables** |
|---|---|---|---|
| PC 1 | 26.1% | Property, plant and equipment<br>Net Property, plant and equipment | Depreciation<br>Amortization |
| PC 2 | 16.4% | Property, plant and equipment<br>Other intangibles | Other financial flows<br>Earnings after tax, Net profits |
| PC 3 | 9.5% | Other short term debt<br>Intangibles | R&D Expenditures<br>Property, plant and equipment |

[18] Given the structure of the dataset, where variables are quite heterogeneous in sample size, a strategy to cope with missing values has been introduced according to the established literature: Schafer (1999) evaluated as likely biased an analysis with more than a missing rate of 5%, while Bennett (2001) established a level of 10% of data missing. Tabachnick and Fidell (2012) stressed instead the role of missing data mechanisms and missing data patterns. In pratice, we have introduced an imputation algorithm based on the mean, considering only those variables that present a number of observations at least close to 85% of the total number (17.000 over 19.852). Before imputation, the sample has been transformed in per worker term and normalized.
[19] All the other components explain less than 8% of the variance of the original features and are quite heterogeneous in the two periods. For example, the fourth principal component is positively related to R&D expenditures (and intangibles) and negatively to other financial flows in the pre-Covid period, while it is strongly positively related only to intangibles in the post-Covid period. The other PCs are related to capital efficiency measures, debts/credits and efficiency in sales (detailed results are available upon request).

*Table 4 – correlations between principal components and features, post-Covid period.*

| (post-Covid) | Var. Explained | Top Positive Correlated Variables | Top Negative Correlated Variables |
|---|---|---|---|
| PC 1 | 27.7% | Property, plant and equipment<br>Net Property, plant and equipment | Depreciation<br>Amortization |
| PC 2 | 12.7% | Property, plant and equipment<br>Cost of Goods sold, Earnings tax | Amortization<br>Net profits, Earnings after tax |
| PC 3 | 11.9% | Accounts receivable | Property, plant and equipment<br>R&D Expenditures |

## 4.3 Cluster Analysis

This section shows the results from the cluster analysis conducted combining principal components and TFP growth. This allow to manage firms' heterogeneity for better representing firms' characteristics and serving as a classifier for in depth studies on the relation between TFP and its determinants.[20]

### 4.3.1 Self-Organizing Maps

Self-organizing maps (SOM) are introduced before clustering to give a better interpretation of clusters and their proximity. These maps allow to visualise the entire dataset in a two dimensional grid made by nodes ordered on the basis of the input variables (principal components and TFP growth estimates). In particular, nodes assume a "U"-shape distribution, where the corners represent respectively maximum and minimum productivity growth nodes, while the center of the distribution is made by highly dense nodes with almost no productivity growth (Figure A.2 in the Appendix).

The output of the SOM models is then used used for performing the k-means algorithm, while a direct application of the k-means on the dataset is used as robustness check. Based on Elbow method and gap statistics five clusters are identified in the pre-Covid period and six in the post-Covid period.[21]

### 4.3.2 Firms groups in the pre-Covid period

According to k-means, SOM nodes are allocated into five different groups, shown in Figure 2. Four of them are distributed according to TFP growth: very low productive growth, low productive growth, medium productive growth, very high productive growth, while the fifth group is more related to specific principal components. The results for the firms' sample are shown in Table 5.

---

[20] K-means clustering algorithm is implemented in such a way that the difference among groups is maximized and the difference within each group is minimized. In other words, all the elements that belong to a specific group are expected to be the most like each other and the most different to the elements of the other groups. This (dis)similarity is measured according to the standard setting that considers Euclidean distance and complete linkage (James, Witten, Hastie, & Tibshirani, 2017).

[21] These two methodologies are slightly different. According to the Elbow method, the number of clusters is selected at the point in which the drop in the WSS is the largest (excluding the first one), while with the Gap statistic the selection is made at the maximum Gap. In our case, local maxima are selected due to a reduction in Gap growth (results available upon request).

*Figure 2 – K-means clustering using SOM nodes (pre-Covid period).*

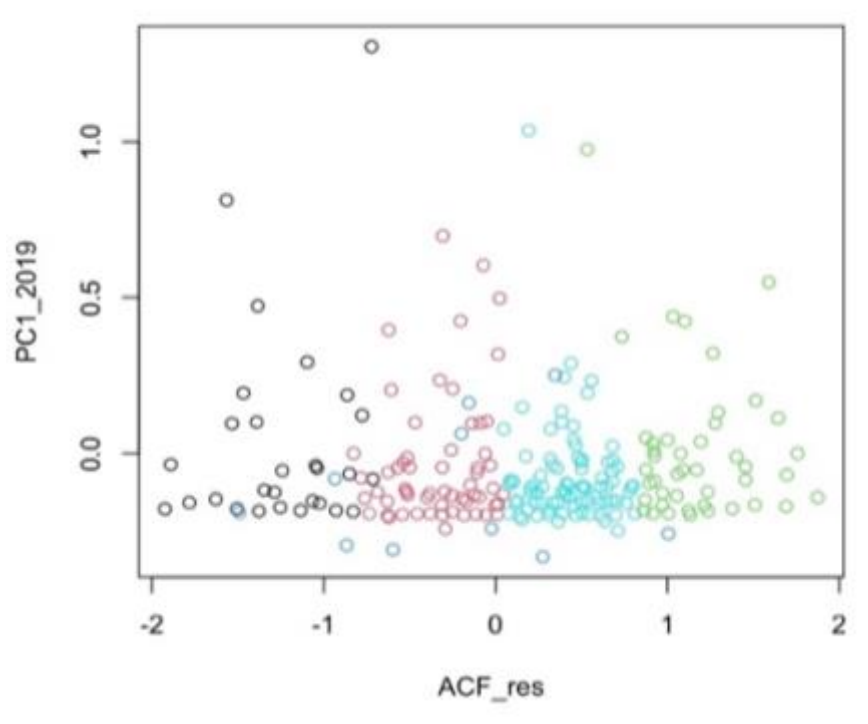

*Table 5 – Groups' characteristics according to PCs, pre-Covid period.*[22]

| Groups | G1 | G2 | G3 | G4 | G5 |
|---|---|---|---|---|---|
| **N** | **1623** | **71** | **18** | **4826** | **2141** |
| **ACF_res** | **-1.02268** | **-0.4624** | **0.0757** | **0.12276** | **1.02226** |
| **PC1** | -0.12946 | -0.2193 | 0.8797 | -0.14369 | -0.11875 |
| **PC2** | -0.03971 | 0.7850 | 0.0154 | -0.04299 | -0.06330 |
| **PC3** | -0.12137 | 0.7429 | 0.0122 | -0.10695 | -0.09738 |

The first group is characterized to have very low values for productivity growth, which means that firms' productivity grows at a slower rate than the average firm. Looking at the most relevant principal components, this group has negative values for PC1 (-0.13) and PC3 (-0.12), and is almost zero for PC2 (-0.04), denoting low efficiency in the use of fixed assets, R&D expenditures and profitability.

Even the second group has low productivity growth, associated with negative values for PC1 (-0.22). However, it relates positively to PC2 (0.79), and PC3 (0.75). In this context, firms' productivity grows faster than that of the first group, but the number is firms is quite limited (71 vs 1623). This is reasonable, considering that these firms have low productivity growth although the level of the principal components related to profitability and to R&D effort (the relation is inverse) is quite large.

At this point, we may say that low productivity growth seems to be associated with low capital efficiency (property, plant, equipment per worker), while the role of R&D activity must be discussed more in detail. Group 2 firms perform better costs and debit/credits management (cost of goods sold and accounts receivable/short term debt per workers) than group 1, but they seem to fail more on average in R&D effort although large profitability (PC4 is positively related to R&D expenditures and it's very low).

Moving up to the third and fourth groups we have slightly positive values for productivity growth (almost the same, firms' productivity grows slightly faster than the average firm), while there are differences in the principal component related to capital efficiency and R&D effort: group 3 has positive values for PC1 (on average 0.88), while group 4 has negative values (on average -0.14). Moreover, group 3 is the only group with positive values for PC1, meaning that it benefits from a stronger role of property, plant and equipment (capital efficiency). Group 4 is instead characterized by negative values for PC3 (-0.11), denoting a positive association with R&D effort (inverse relation). Even in this case, group 3 is largely smaller than group 4 (18 vs 4826 observations).

[22] Values taken by the other principal components are available upon request.

Finally, group 5 is characterized by the largest average TFP growth, and slightly negative values for the first three principal components. This group may take advantage from higher net profits (inverse relation to PC2) and it shows important contributions of the principal components related to R&D expenditures and intangibles, interpretable as R&D effort (PC3) and R&D outcome (PC4).

### 4.3.3 Firms groups in the post-Covid period

Even in the post-Covid period we observe at least four groups related to the intensity of TFP growth, as shown in Figure 3. The results for the firms' sample are shown in Table 6.

However, the main differences here are represented by the presence of six groups instead of five and by the different combinations of the principal components. For these reasons, we cannot compare directly all the clusters in the two periods, but we may look at the determinants that emerge from each analysis.

***Figure 3 – K-means clustering using SOM nodes (post-Covid period).***

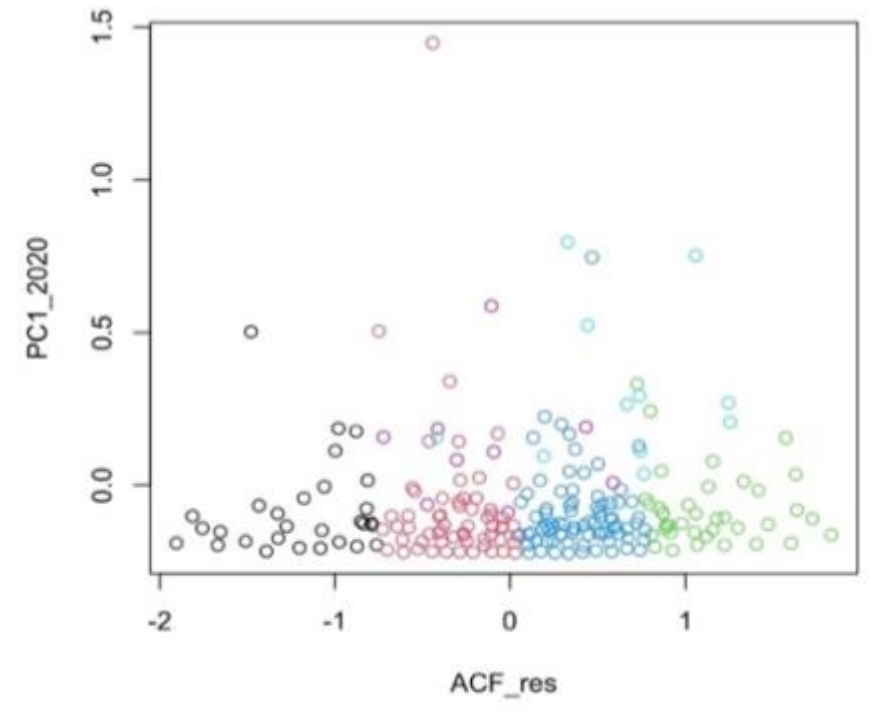

***Table 6 – Groups' characteristics according to PCs, post-Covid period.***[23]

| Groups | G1 | G2 | G3 | G4 | G5 | G6 |
|---|---|---|---|---|---|---|
| **N** | **3073** | **145** | **27** | **18** | **4518** | **13** |
| **ACF_res** | **-0.59862** | **-0.04649** | **0.1289** | **0.234** | **0.5984** | **1.055** |
| **PC1** | -0.16036 | 0.00162 | 1.0284 | 0.553 | -0.1505 | 0.334 |
| **PC2** | 0.04671 | 0.24017 | 0.1290 | -0.789 | 0.0134 | -0.680 |
| **PC3** | -0.10376 | 0.55838 | -1.2222 | 0.349 | -0.0803 | 0.275 |

[23] Values taken by the other principal components are available upon request.

Groups 1 and 2 are again associated with negative values of ACF residuals (TFP grows slower than the average), non-positive values for PC1 (-0.6 and 0), while they both have positive values for PC2 (0.05 and 0.24). It seems that on average these groups experienced an increase in the average TFP growth, probably due to a larger number of firms classified in group 1 in the post-Covid period, and a convergence in the relation with the second principal component, associated negatively to profitability (net profit per workers). Even the values for the third principal component are quite similar. Note that group 2 almost doubled in size (145 vs 71), leading to an increase in the heterogeneity of that group.
Groups 3 and 4 stay instead on their medium productivity growth (TFP grows only slightly faster than the average). Both groups are positively related to PC1 (1.03, 0.55), but group 3 is positively related to PC2 and negatively to PC3 while group 4 is the reverse. Capital efficiency (property, plant, equipment per worker) plays again a major role in both groups, with opposite dynamics in the other components. In particular, Group 3 is characterized by high coefficients associated with R&D effort, but not with R&D outcome, while Group 4 seems to have low levels of R&D efforts but potential outcome.
Finally, group 5 in the post-Covid period is almost peculiar to the corresponding group in the pre-Covid period, but it doubles the number of observations (4518 from the 2141 of the pre-Covid period). This implies a drop in the average TFP of the group, which remains at a high level, while the other components seem to be less affected. Group 6 is, instead, the one with the best performance in terms of TFP growth although the scarce numerosity (just 13 observations).

### 4.3.4 More on clusters' characteristics: main sectors and taxonomies

An important point of the cluster analysis consists of identifying groups that are better representative of firm productive behavior than the original sectors as well as other classifications known in the literature. In the pre-Covid period, the most representative sector was Industrial, Electric & Electronic Machinery (between 10 and 15% in all the five groups), followed by Chemicals, Petroleum, Rubber & Plastic (about 10% in all the groups). At this point, the structure of the clusters seems to be quite similar.
However, things become different when looking at the third most important sector. Groups 1, 2, 4 and 5 are characterized by a relevant share of Business services (about 7%), while group 3 firms share the frequency of this sector with many others, representing a quite heterogeneous group. Moreover, high-productive growth groups present larger shares of the Wholesale sector with respect to the others.
The findings for the clusters in the post-Covid period are not so different but consist of some changes. Industrial, Electric & Electronic Machinery (between 10 and 25%) is confirmed to be the most representative sector in all groups, except for group 4 where it is absent at all. Even Chemicals, Petroleum, Rubber & Plastic (between 10 and 12%) remain the second most representative sector for all the clusters with the exception of group 6. In this group, the property services sector is the one that emerges with a high frequency. Groups 3 and 4 are (again) more uniform in their distributions, while the others (1, 2 and 5) remain the most heterogeneous.

Looking at the classification of sectors based on the Schumpeterian patterns of innovation[24] for all the clusters and in both periods, Schumpeter mark I is the most representative with more than the 60% of the observations. This result is reasonable, since Schumpeter mark II firms are mostly characterized to be persistent innovators, while Schumpeter mark I firms are more heterogeneous.

Things are slightly different when looking at the distributions of the firm's sectors according to the Pavitt taxonomy.[25] In almost all the clusters specialised suppliers are the most relevant category, followed by the suppliers dominated. It's interesting that suppliers dominated firms prevail in group 3 pre-Covid and in group 4 post-Covid, both characterized by a slightly positive level of TFP growth. Science-based is about 10% with a larger presence in cluster 2 pre-Covid and cluster 4 post-Covid.

In general, it seems that original sectors and classic taxonomies cannot express differences in terms of productivity growth, while the new clusters can. Despite that, there are some groups that are quite similar in terms of average TFP growth. It's thus necessary to look more in depth at clusters' characteristics to better understand the differences between these groups.

### 4.3.5 More on clusters' characteristics: clusters' dissimilarity

In this paragraph we aim at first to verify if the average productivity growth of each cluster is statistically different from the ones of the other clusters. Average productivity estimates are tested to be statistically different two at a time using the student-t test for the comparison of two sample means.[26]

In the pre-Covid period, only the average TFP growth of groups 3 and 4 is not statistically different: both medium productive groups share the same TFP growth. However, group 3 is characterized to be assigned to unique values of the principal components more than on a particular TFP growth level.

Looking at the post-Covid period, the average TFP growth of groups 3 and 4 (associated with medium productivity growth) is again not statistically different. Moreover, in this case also the average TFP growth of groups 2 and 3 is quite similar.

Mean values of the original variables within each cluster (both in absolute values and per worker) are now considered to identify the most prominent directions of the principal components and to understand better which variables explain differences among groups and, in particular, between the groups that share the same (average) value for TFP growth.

---

[24] Based on these classification, original sectors are identified as Schumpeter mark I, Schumpeter mark II or Public service, according to their characteristics (Malerba & Orsenigo, 1995).

[25] According to Pavitt, firms are classified as supplier dominated, specialized suppliers, scale intensive or science based (Pavitt, 1984). Moreover, a category for public services has been added.

[26] The results of the student-t tests are shown in Tables A.5 and A.6 in the Appendix for both the pre and the post-Covid periods.

The groups associated with positive values of the first principal component are group 3 pre-Covid and groups 3, 4 and 6 post-Covid. As reported in Tables A.7-A.10 in the appendix (related to the most relevant variables within each cluster both in absolute and in per workers term), these groups have the lowest absolute and relative values for property, plant & equipment measures, the first loadings of PC1. Firms belonging to these clusters are not particularly efficient in the use of fixed assets in the long run and they just reach medium levels of productivity growth in both periods. The only exception is represented by cluster 6 post-Covid (this may be an effect of the pandemic on particular firms).

Other important considerations are related to the dimensions of the second principal component, which is strongly negative for groups 5 pre-Covid and 4 and 6 post-Covid. Again, as shown in Tables A.7-A.10, these groups collected the largest amount of net profits among all the groups in absolute values in both periods. However, looking at the profitability (net profit per worker) only group 5 pre-Covid confirms this result. Groups 4 and 6 of the post-Covid period are, instead, not doing particularly well in profitability, but they realize a good performance in cost management (cost of goods sold per worker is the lowest among all groups).

The fact that all these groups reach high levels of productivity growth proves that both profitability and efficiency in cost management are important determinants of TFP growth. The fact that the second one prevails after the outbreak of Covid pandemic may represent a direct consequence of the reaction of the firms to the pandemic shock.

While the dimensions of the third principal components are quite heterogeneous, we can look at the values of R&D expenditures and intangibles in both absolute and relative terms.

R&D expenditures are larger in groups 1, 4 and 5 pre-Covid and in groups 1 and 5 post-Covid. All these groups, indeed, are related negatively to the third principal component in the pre-Covid setting (and positively to the fifth principal component in the post-Covid setting). Per workers values confirm this result only in the post-Covid setting, while in the pre-Covid the distribution of R&D expenditures is more homogeneous. The interpretation is quite intuitive: in a situation of economic stability firms sustain on average the same effort in the R&D activity, while the pandemic shock affected firms in different ways changing their choices.

Looking at the values of the output of the R&D activity, as proxied by the intangibles per workers variable (R&D intensity), groups 4 and 5 pre-Covid and groups 1 and 5 post-Covid are the ones with the highest values. Not surprisingly in both periods group 5 firms, the ones with high productivity growth, conduct the R&D activity efficiently. Less clear is the result of group 1 post-Covid: it seems that TFP growth has slowed for this group despite a good efficiency of the R&D activity.

### 4.3.6 Changes in clusters membership (pre- and post-Covid)

Since the numerosity of the clusters changes drastically in the two periods, it doesn't make any sense to compare the absolute values of the average TFP growth without considering how many firms have been classified differently in the two periods. Moreover, analyzing the changes in the cluster membership between the two periods allows to evaluate better how the Covid pandemic may have affected firms.
Table A.11 in the Appendix shows that the largest changes are coming from the largest groups: firms seem to move to higher productive growth groups, meaning that the Covid pandemic has worsened the economic conditions of some firms only, while the others have potentially improved theirs. Furthermore, the large size of group 5 in the post-Covid period can be explained by the different classifications of group 1 and 4 firms in the pre-Covid period. Although the average TFP growth of this group remains one of the highest, it drops significantly if we consider group 5 of the pre-Covid period as a benchmark. On the contrary, the reclassification of several firms from this last group to group 1 in the post-Covid period makes the average TFP growth of this group to increase slightly. A trend of convergence in the average TFP growth is therefore observed because of the Covid pandemic.
Interestingly, the size of the movements of firms between groups 1 and 5 in the two periods is almost the same. All these groups show efficiency in conducting their R&D activity. We can just conclude that the pandemic affected more the other TFP determinants than the R&D effort/efficiency.

## 4.4 Principal Component Regression

If the presence of different clusters allows us to deal with heterogeneity in firm characteristics, the point now is to consider how and if principal components may be good proxies in explaining the relation between TFP growth estimates and its determinants. Moreover, thanks to the results of the cluster analysis it is possible to check if this relation changes within each different group.
The way in which this is implemented is through the principal component regression that implies the presence of a linear relationship between TFP and principal components. Figures 4 and 5 show the results of the PCR in the entire sample of firms for both periods using the estimate of the productivity growth (ACF) as the response variable, and up to eight principal components as explanatory variables.
Looking at the estimates in the pre-Covid period, significant coefficients are associated with seven principal components. In particular, looking at the most relevant components, PC3 shows a positive coefficient (0.21), while PC1 and PC2 have negative (-0.52, -0.41). These denote negative relationships between TFP growth and capital efficiency, and positive relationships between TFP growth and firm's profitability (inverse relation), and R&D effort/efficiency.
Although the variance explained by the principal components is different, results seem to be quite similar looking at the post-Covid period. In particular, significant coefficients are associated (again) with seven principal components, PC3 has positive coefficient (0.26), while PC1 and PC2 negative (-0.57, -0.21).

***Figure 4 – PC regression results for both periods without controls (full sample).***

| (ACF) | pre-Covid period | | | post-Covid period | | |
|---|---|---|---|---|---|---|
| | estimate | p-value | | estimate | p-value | |
| PC1 | -0.5230 | 0.00 | *** | -0.5650 | 0.00 | *** |
| PC2 | -0.4141 | 0.00 | *** | -0.2117 | 0.0049 | ** |
| PC3 | 0.2125 | 0.0024 | ** | 0.2550 | 0.00 | *** |
| PC4 | 0.1649 | 0.0156 | * | -0.5044 | 0.00 | *** |
| PC5 | 0.3998 | 0.00 | *** | -0.0144 | 0.8499 | |
| PC6 | 0.2644 | 0.00 | ** | -0.9647 | 0.00 | *** |
| PC7 | 0.0553 | 0.4559 | | -0.1409 | 0.0453 | * |
| PC8 | -0.9173 | 0.00 | *** | 0.3719 | 0.00 | *** |

***Figure 5 – PC regression results for both periods with controls (full sample).***

| (ACF) | pre-Covid period | | | post-Covid period | | |
|---|---|---|---|---|---|---|
| | estimate | p-value | | estimate | p-value | |
| PC1 | - 0.2148 | 0.002 | ** | - 0.18762 | 0.01236 | * |
| PC2 | - 0.4383 | < 0.002 | *** | - 0.30444 | < 0.002 | *** |
| PC3 | 0.3385 | < 0.002 | *** | 0.29996 | < 0.002 | *** |
| PC4 | 0.1053 | 0.1246 | | - 0.45836 | < 0.002 | *** |
| PC5 | 0.3741 | < 0.002 | *** | 0.00435 | 0.95438 | |
| PC6 | 0.2406 | 0.0028 | ** | - 0.79532 | < 0.002 | *** |
| PC7 | 0.0185 | 0.8033 | | - 0.24969 | < 0.002 | *** |
| PC8 | - 0.7642 | < 0.002 | *** | 0.36492 | < 0.002 | *** |
| *country_d* | | *yes* | | | *yes* | |
| *sector_d* | | *yes* | | | *yes* | |

Considering the principal component regressions conducted by restricting the sample to the observations of each cluster separately in both periods with and without controls (results available upon request), the situation is quite heterogeneous. In general, introducing controls is not always possible due to the lack of data on specific clusters, but when it is possible it affects slightly the within sample results.

In the pre-Covid period, except for cluster 2, all the others have significant coefficients that link the principal components with TFP growth. The situation is slightly worse in the post-Covid period when only clusters 1 and 5 have significant coefficients. This confirms the results of the previous steps of the analysis: groups that are efficient in conducting the R&D activity show that a relation between PCs and TFP growth is true even after the Covid shock.

#### 4.4.1 LASSO Regression

Finally, Lasso regression is implemented to obtain significant estimates of the coefficients of the original variables that are related to TFP growth. As shown in Table A.12 in the Appendix, the significant coefficients of the Lasso are equivalent to the main loadings of the principal components: property, plant, equipment (PC1), and cost of goods sold (PC2).

The presence of property, plant and equipment coefficient is particularly relevant since confirms the negative coefficient associated with PC1 in the principal component regression. The main interpretation of this result is that the efficiency in the use of fixed assets, in the long run, may reduce productivity growth in the short run due to the limited amounts of available resources. This is confirmed also by a particular role of debt which emerges for the principal component regression. Firms with additional resources present indeed larger values of productivity growth.

## 5. MAIN RESULTS AND CONCLUSIONS

This paper has investigated the relation between total factor productivity and its determinants using a bottom-up approach that combines principal component analysis, self-organizing maps, clustering, and principal component regression. By relying on unsupervised learning methods, the analysis moves from conventional sector-based approaches and instead exploits the underlying structure of firm-level data to characterize better firms' heterogeneity. This perspective is consistent with the economic literature that investigate the importance of micro-level variation in explaining aggregate productivity performance.

The empirical findings indicate that several determinants - capital efficiency, profitability and cost management, credit/debts measures, and R&D effort/efficiency - represent the most relevant firms' characteristics in explaining differences in TFP growth in both the pre-Covid and post-Covid period.

In addition, the paper has provided a further classification, different from the original sectors and other taxonomies present in the literature: the cluster analysis allowed us to identify at least five clusters according to the average levels of TFP growth from very low to very high. In particular, five clusters have been found in the pre-Covid period and six in the post-Covid period.

Some clusters are characterized by similar features as measured by the principal components. The first principal component, related to capital efficiency, is positive for group 3 pre-Covid and groups 3, 4 and 6 post-Covid, while it is negative for all the other groups. The second principal component, associated positively with the efficiency in cost management and negatively with profitability is always non-positive for all the groups in the pre-Covid period and it is positive only for groups 2 and 3 in the post-Covid period. Moreover, the clustering analysis shows that groups with similar TFP growth (like groups 3 in pre-Covid period and groups 3, 4 and 6 in the post-Covid period) may differ substantially in their structural profiles, suggesting the presence of resource misallocation and adjustment mechanisms. Other groups, especially the ones with the largest samples (groups 1 and 4 in the pre-Covid period and groups 1 and 5 in the post-Covid period), relate in the same manner with the TFP determinants. This proves that the difference in TFP growth levels is mainly given by the way in which firms can better exploit their opportunities and make decisions (productivity decisions, investment decisions, managerial practice and organizational framework all have impacted output levels and thus the way in which each firm uses efficiently inputs). The fact that the TFP growth is lower in some groups comes not only from profitability and efficiency in the use of resources, which are observable but also by the outcome of the decision made by the firms.

Not surprisingly, in both periods there are several principal components that refer to firms' R&D effort/efficiency (R&D expenditures or intangibles per worker). The difference observed with respect to these determinants between the two periods confirms the impact of the Covid pandemic in increasing uncertainty and in disrupting the R&D activity conducted by the firms. Moreover, the Covid shock appears to have altered the distribution of firms across clusters, reducing the dispersion of TFP and increasing the importance of capital and cost related determinants.

Finally, the principal component regression has confirmed a statistically robust association between TFP growth and the main determinants, while the within-cluster regressions reveal that this relationship is not consistent across all the clusters. LASSO results strengthen the relevance of the variables related to capital intensity, cost structure, intangible assets, and sales performance, reinforcing the robustness of the findings.

Our methodology, and the resulting taxonomy based on largely available accounting data, can also be valuable for policy. Indeed, classifying firms, and identifying some characteristics typically associated with each group, allows policymakers to pinpoint where resources and interventions are most needed, whether it is in boosting lagging firms or supporting high-performing ones. For instance, high-productivity firms may benefit from policies that encourage cutting-edge research and development, while lower-productivity firms might need assistance with technology adoption or skills training. By understanding the distinct characteristics and challenges of each group, policies can be tailored to foster innovation, enhance competitiveness, and ultimately promote economic growth. In addition, the proposed approach can help in monitoring policy outcomes and adjusting strategies as needed. Indeed, the fact that the identified clusters do not align with conventional sectoral classifications indicates that industrial policy may benefit from a micro-targeted approach, oriented toward groups of firms sharing structural characteristics rather than broad sectoral categories.

Overall, the findings demonstrate that a bottom-up, unsupervised learning framework provides valuable insights into the complex relationship between firm characteristics and TFP growth. Future research could extend this analysis by incorporating longer time horizons, additional dimensions such as managerial quality or human capital, and causal identification strategies to further investigate the mechanisms underlying productivity growth.

# APPENDIX

***Table A.1 – Descriptive statistics for the balance sheet (USD), averages 2015-2019.***

| Variable | N | Mean | Pctl. 85 | Max | Std. Dev. |
|---|---|---|---|---|---|
| **assets_current** | 19852 | 792949.70 | 590540.38 | 196541906.32 | 4944837.03 |
| **inventory** | 17412 | 163675.80 | 119623.82 | 66442324.95 | 1143767.68 |
| **accounts_receivable_net** | 19415 | 185304.88 | 147655.51 | 77314091.41 | 1096161.51 |
| **accounts_receivable** | 18370 | 194685.68 | 159638.91 | 77675039.09 | 1130381.33 |
| **assets_other** | 19852 | 475614.07 | 291247.18 | 192678588.58 | 3659333.10 |
| **assets_current_other** | 19439 | 195712.67 | 64584.36 | 140614302.20 | 2181400.50 |
| **cash_short_term_investment** | 19846 | 248943.24 | 179369.07 | 121959800 | 1906118.49 |
| **cash** | 19843 | 189148.85 | 149513.83 | 55929431.13 | 1093198.82 |
| **property_plant_equipment_net** | 19726 | 622546.05 | 305044.18 | 258285075.03 | 4956379.46 |
| **property_plant_equipment_oth_tot** | 17870 | 344011.42 | 84421.06 | 337561000 | 5269791.33 |
| **property_plant_equipment_oth_net** | 17808 | 173668.89 | 45457.34 | 158245900.90 | 2243213.73 |
| **intangibles** | 17056 | 430470.50 | 125855.85 | 162182600 | 3309060.29 |
| **intangibles_other** | 17019 | 185106.03 | 51864.34 | 59808256.69 | 1550084.96 |
| **assets_fixed_other** | 19313 | 404439.62 | 165193.91 | 223296400 | 3782568.43 |
| **assets_long_term_other** | 18690 | 183164.37 | 76935.45 | 149199496.93 | 1923786.37 |
| **assets_tot** | 19852 | 2173104.11 | 1440536.25 | 485386918.11 | 13187883.34 |
| **liabilities_current** | 19852 | 609855.50 | 384360.97 | 184420672.18 | 4085172.55 |
| **creditors_trade** | 19459 | 176718.68 | 110655.93 | 46854681.11 | 1202840.46 |
| **liabilities_other** | 19852 | 314870.91 | 159181.73 | 99020071.45 | 2243086.86 |
| **debt_short_term_other** | 18785 | 35655.12 | 6812.76 | 34236247.49 | 416005.52 |
| **liabilities_current_other** | 19640 | 187110.06 | 83994.31 | 54231000 | 1398132.80 |
| **liabilities_non_current** | 19838 | 704631.59 | 314841.02 | 203026702.30 | 4828030.62 |
| **debt_interest_long_term** | 17944 | 519180.75 | 271769.31 | 102142069.66 | 3159108.34 |
| **liabilities_non_current_other** | 19785 | 238255.55 | 83638.02 | 144147161.72 | 2195785.14 |
| **liabilities_debt_total** | 19852 | 1313678.14 | 746957.25 | 365763296.88 | 8452190.35 |
| **shareholders_equity_tot** | 19107 | 899493.78 | 671735.36 | 205808202.60 | 5541464.89 |
| **shareholders_equity_net** | 17520 | 845757.51 | 632871.25 | 191196900.78 | 5268409.70 |
| **EV** | 16484 | 2468753.78 | 1921643.8 | 854929969.13 | 15487301.80 |

*Table A.2 – Descriptive statistics for the balance sheet (USD), year 2020.*

| Variable | N | Mean | Pctl. 85 | Max | Std. Dev. |
|---|---|---|---|---|---|
| assets_current | 19848 | 1005710.342 | 738982.642 | 291501619.46 | 6701250.699 |
| inventory | 16827 | 195942.095 | 143142.91 | 162809457.25 | 1742525.34 |
| accounts_receivable_net | 19157 | 200029.071 | 158885.279 | 81540914.01 | 1176973.731 |
| accounts_receivable | 17147 | 209296.03 | 176311.484 | 81826243.73 | 1221999.848 |
| assets_other | 19846 | 650652.493 | 397522.97 | 284406637.38 | 5257199.8 |
| assets_current_other | 18940 | 280444.98 | 82905.309 | 231083405.54 | 3535012.452 |
| cash_short_term_investment | 19829 | 343666.558 | 250668.615 | 136694000 | 2480218.224 |
| cash | 19827 | 272960.703 | 213448.556 | 49443907.12 | 1531612.476 |
| property_plant_equipment_net | 19659 | 693134.804 | 361083.87 | 233631000 | 5285344.583 |
| property_plant_equipment_oth_tot | 17192 | 408371.353 | 106156.111 | 402427107.91 | 6208700.484 |
| property_plant_equipment_oth_net | 17023 | 187068.683 | 48338.675 | 160724871.55 | 2325285.654 |
| intangibles | 16205 | 552493.286 | 166391.028 | 162498000 | 4224432.001 |
| intangibles_other | 16156 | 242106.515 | 65770.461 | 120558824.87 | 2234783.319 |
| assets_fixed_other | 19082 | 559624.165 | 244903.473 | 240112100.56 | 4835906.506 |
| assets_long_term_other | 18274 | 297844.594 | 132798.487 | 202525803.51 | 2894917.001 |
| assets_tot | 19852 | 2680904.584 | 1823887.974 | 610008241.14 | 16107664.753 |
| liabilities_current | 19852 | 759549.754 | 468204.778 | 230646673.55 | 5379639.779 |
| creditors_trade | 19207 | 204921.461 | 120864.236 | 95137642.22 | 1595024.091 |
| liabilities_other | 19850 | 419187.703 | 211749.117 | 158435012.56 | 3150677.028 |
| debt_short_term_other | 17905 | 56326.889 | 11401.597 | 52838895.83 | 611292.605 |
| liabilities_current_other | 19281 | 210391.342 | 99667.37 | 72800212.69 | 1698315.906 |
| liabilities_non_current | 19833 | 909471.914 | 417481.696 | 249004216.9 | 6231045.774 |
| debt_interest_long_term | 16801 | 729272.954 | 412627.131 | 148827511.43 | 4346929.789 |
| liabilities_non_current_other | 19782 | 292090.962 | 102372.505 | 176931766.69 | 2691093.881 |
| liabilities_debt_total | 19852 | 1667799.426 | 955710.389 | 451978712.06 | 10836165.681 |
| shareholders_equity_tot | 18864 | 1080680.103 | 841071.705 | 222544001 | 6355806.941 |
| shareholders_equity_net | 17249 | 1034871.104 | 811176.517 | 222543324.78 | 6123186.775 |
| EV | 16459 | 3542317.243 | 2513585.8 | 2029991303 | 29594205.604 |

***Table A.3 - Descriptive statistics for income statement (USD), averages 2015-2019.***

| Variable | N | Mean | Pctl. 85 | Max | Std. Dev. |
|---|---|---|---|---|---|
| **revenues_tot** | 19848 | 1334329.164 | 1013771.835 | 307507800 | 8019940.901 |
| **sales_gross** | 19848 | 1329769.011 | 1012482.864 | 307397200 | 7981159.423 |
| **goods_sold_cost** | 18871 | -849331.691 | -2321.911 | 0 | 5561572.022 |
| **rd_expenses** | 19815 | -35165.552 | 0 | 0 | 394565.936 |
| **items_operating_other** | 19795 | -285266.955 | -1412.935 | 0 | 1703017.783 |
| **depreciation_amortization** | 19790 | -91246.637 | -162.587 | 0 | 684322.83 |
| **depreciation** | 19783 | -71408.441 | -110.57 | 0 | 600337.618 |
| **amortization** | 19436 | -20549.491 | 0 | 0 | 187819.617 |
| **financial_expenses** | 19374 | -23711.743 | -23.811 | 0 | 146641.961 |
| **earnings_tax** | 17170 | -34717.898 | -39.421 | 0 | 236845.594 |
| **dividends** | 18811 | -165960.836 | 0 | 0 | 15704188.341 |
| **sales_net** | 19852 | 1323875.884 | 1004422.969 | 307397200 | 7965886.197 |
| **EBITDA** | 19851 | 213567.195 | 131946.684 | 76115200 | 1458440.817 |
| **operating_income** | 19852 | 122658.18 | 80521.279 | 65485200 | 920918.679 |
| **EBIT** | 19852 | 116624.674 | 77420.167 | 65485200 | 904809.707 |
| **financial_revenue** | 18813 | 6754.058 | 2978.944 | 6205923.582 | 81825.786 |
| **financial_result** | 19750 | -16946.048 | 54.305 | 5654067.692 | 127294.892 |
| **financial_flow_other** | 19041 | 12365.438 | 4311.439 | 8863200 | 170651.019 |
| **earnings_before_tax** | 19852 | 111642.622 | 71778.515 | 67323200 | 925810.656 |
| **earnings_after_tax** | 19852 | 82793.432 | 53444.417 | 52743800 | 729604.601 |
| **interest_minority** | 18519 | -5092.249 | 1.211 | 1615461.448 | 64888.642 |
| **profit_net** | 19852 | 78893.377 | 51899.466 | 52443800 | 706729.281 |
| **cash_flow** | 19712 | 171391.883 | 102536.679 | 63073800 | 1236795.444 |
| **earnings_retained** | 19701 | 469587.681 | 261127.74 | 415561800 | 5300063.53 |
| **share_reserves** | 18873 | 110452.067 | 50650.84 | 103486200 | 1660376.479 |
| **assets_net** | 19852 | 866189.242 | 633816.699 | 205808202.596 | 5466856.635 |
| **debt_net** | 19852 | 373766.284 | 178746.635 | 148583920.194 | 3296138.193 |

*Table A.4 - Descriptive statistics for income statement (USD), year 2020.*

| Variable | N | Mean | Pctl. 85 | Max | Std. Dev. |
|---|---|---|---|---|---|
| **revenues_tot** | 19845 | 1421365.873 | 1081463.077 | 386064000 | 8347597.528 |
| **sales_gross** | 19842 | 1411830.724 | 1077775.089 | 386064000 | 8275956.021 |
| **goods_sold_cost** | 18551 | -890829.702 | -2165.184 | 0 | 5494320.822 |
| **rd_expenses** | 19818 | -44653.888 | 0 | 0 | 566226.442 |
| **items_operating_other** | 19770 | -319305.92 | -1453.395 | 0 | 2018319.61 |
| **depreciation_amortization** | 19811 | -123695.264 | -182.866 | 0 | 965889.892 |
| **depreciation** | 19804 | -93625.579 | -128.909 | 0 | 793351.366 |
| **amortization** | 19378 | -30779 | 0 | 0 | 346556.588 |
| **financial_expenses** | 18911 | -28435.023 | -25.287 | 0 | 166547.733 |
| **earnings_tax** | 15579 | -39149.031 | -23.209 | 0 | 270425.844 |
| **dividends** | 18148 | -402207.423 | 0 | 0 | 48269842.221 |
| **sales_net** | 19852 | 1405923.154 | 1073072.984 | 386064000 | 8260824.688 |
| **EBITDA** | 19851 | 229113.556 | 142224.345 | 81154057.632 | 1768300.941 |
| **operating_income** | 19852 | 105690.553 | 75750.273 | 73272612.127 | 1274271.815 |
| **EBIT** | 19852 | 98731.04 | 71784.871 | 66288000 | 1216161.062 |
| **financial_revenue** | 17047 | 7355.74 | 2963.627 | 10172248.278 | 108899.182 |
| **financial_result** | 19525 | -21117.261 | 40.339 | 9489962.403 | 164363.286 |
| **financial_flow_other** | 18440 | 10438.733 | 5190.285 | 18954082.654 | 309063.862 |
| **earnings_before_tax** | 19852 | 87657.884 | 66089.781 | 67091000 | 1279450.309 |
| **earnings_after_tax** | 19852 | 61694.767 | 49939.58 | 57411000 | 1073979.571 |
| **interest_minority** | 18452 | -5279.648 | 249 | 2043000 | 83385.96 |
| **profit_net** | 19852 | 58599.934 | 47867.115 | 57411000 | 1072749.431 |
| **cash_flow** | 19508 | 185316.346 | 113164.415 | 68467000 | 1535836.518 |
| **earnings_retained** | 19496 | 531075.3 | 327700.906 | 383943000 | 5629810.37 |
| **share_reserves** | 18452 | 125803.899 | 52707.319 | 99205000 | 2276457.06 |
| **assets_net** | 19852 | 1020787.46 | 767824.153 | 222544001 | 6232143.518 |
| **debt_net** | 19848 | 466958.901 | 218135.351 | 199390437.597 | 4322852.209 |

*Table A.5 – Student-t tests for the comparison of average TFPs (pre-Covid).*

| Groups | G1 | G2 | G3 | G4 | G5 |
|---|---|---|---|---|---|
| **G1** | - | < 2.2e-16 | < 2.2e-16 | < 2.2e-16 | < 2.2e-16 |
| **G2** | < 2.2e-16 | - | 0.001 | < 2.2e-16 | < 2.2e-16 |
| **G3** | < 2.2e-16 | 0.001 | - | 0.6 | < 2.2e-16 |
| **G4** | < 2.2e-16 | < 2.2e-16 | 0.6 | - | < 2.2e-16 |
| **G5** | < 2.2e-16 | < 2.2e-16 | < 2.2e-16 | < 2.2e-16 | - |

*Table A.6 – Student-t tests for the comparison of average TFPs (post-Covid).*

| Groups | G1 | G2 | G3 | G4 | G5 | G6 |
|---|---|---|---|---|---|---|
| **G1** | - | < 2.2e-16 | 0.0000005 | < 2.2e-16 | < 2.2e-16 | < 2.2e-16 |
| **G2** | < 2.2e-16 | - | 0.2 | 0.03 | < 2.2e-16 | < 2.2e-16 |
| **G3** | 0.0000005 | 0.2 | - | 0.6 | 0.00007 | 0.000004 |
| **G4** | < 2.2e-16 | 0.03 | 0.6 | - | 0.0003 | 0.000002 |
| **G5** | < 2.2e-16 | < 2.2e-16 | 0.00007 | 0.0003 | - | < 2.2e-8 |
| **G6** | < 2.2e-16 | < 2.2e-16 | 0.000004 | 0.000002 | < 2.2e-8 | - |

*Table A.7 – Descriptive statistics for the most relevant variables (USD), pre-Covid.*

| Variables | Group 1 | Group 2 | Group 3 | Group 4 | Group 5 |
|---|---|---|---|---|---|
| **goods_final** | 103373.62 | 91528.70 | 172069.41 | 103476.71 | 96608.57 |
| **workers** | 4534.92 | 3137.90 | 2856.25 | 4385.91 | 4818.37 |
| **assets_fixed** | 1425784.63 | 1425784.63 | 1967017.79 | 1180294.73 | 1901181.73 |
| **investment_ short_term** | 102993.56 | 363050.86 | 330019.72 | 187340.23 | 123304.33 |
| **investments** | 68405.87 | 46060.40 | 86284.07 | 92655.66 | 196576.68 |
| **goods_interm** | 46985.44 | 17766.88 | 17766.88 | 68384.26 | 80437.02 |
| **property_pla-nt_equip._net** | 651688.07 | 1356481.476 | 389480.3407 | 657801.079 | 831673.05 |
| **depreciation_ amortization** | -93269.32 | -183518.686 | -60423.9217 | -88964.676 | -127692.91 |
| **goods_sold_ cost** | -876893.06 | -1061455.692 | -1464904.074 | -876259.862 | -1026193.29 |
| **profit_net** | 72495.45 | 14067.623 | 64738.1136 | 76948.006 | 126477.24 |
| **rd_expenses** | -35004.85 | -9688.007 | -17186.0534 | -38620.339 | -42848.74 |
| **intangibles** | 450364.01 | 208745.042 | 739564.4886 | 409177.116 | 619807.06 |

***Table A.8 – Descriptive statistics for the most relevant per workers variables (USD), pre-Covid.***

| Variables | Group 1 | Group 2 | Group 3 | Group 4 | Group 5 |
|---|---|---|---|---|---|
| **property_plant_equip._net** | 372.20 | 160.610 | 157.1796 | 495.519 | 365.43 |
| **depreciation_amortization** | -38.92 | -21.460 | -13.1795 | -55.030 | -55.77 |
| **goods_sold_cost** | -271.05 | -322.177 | -328.7534 | -508.797 | -270.16 |
| **profit_net** | 12.91 | 2.190 | -47.4574 | 46.468 | 50.24 |
| **rd_expenses** | -22.16 | -20.792 | -70.6737 | -25.011 | -19.37 |
| **intangibles** | 102.97 | 72.849 | 79.4992 | 198.349 | 318.78 |

***Table A.9 – Descriptive statistics for the most relevant variables (USD), post-Covid.***

| Variables | Group 1 | Group 2 | Group 3 | Group 4 | Group 5 | Group 6 |
|---|---|---|---|---|---|---|
| **goods_final** | 122595.83 | 81091.72 | 33089.83 | 299324.96 | 110919.11 | 251339.53 |
| **workers** | 5075.83 | 3369.17 | 3501.23 | 4924.61 | 4625.79 | 5550.30 |
| **assets_fixed** | 1888225.83 | 1614913.02 | 790161.11 | 496146.23 | 1925600.51 | 874602.83 |
| **investment_short_term** | 355331.48 | 51054.71 | 43870.25 | 57061.75 | 160978.54 | 36687.93 |
| **investments** | 173185.88 | 64565.37 | 17357.24 | 9623.34 | 110224.77 | 18910.38 |
| **goods_interm** | 151376.51 | 46891.70 | 24636.83 | 19333.18 | 64002.01 | 9006.5 |
| **property_pla-nt_equip._net** | 813378.799 | 669636.582 | 201982.544 | 165187.768 | 826625.095 | 296963.70 |
| **depreciation_amortization** | -146609.65 | -99081.172 | -28558.596 | -50758.483 | -138340.53 | -129506.24 |
| **goods_sold_cost** | -1113886.7 | -412512.85 | -372515.53 | -1143791.6 | -940694.48 | -633860.18 |
| **profit_net** | 81026.185 | 57684.9158 | 10315.977 | 89939.552 | 42341.395 | 167585.26 |
| **rd_expenses** | -65733.723 | -27341.64 | -9611.107 | -10398.555 | -43337.440 | -9870.12 |
| **intangibles** | 602863.818 | 656485.994 | 188960.232 | 198736.716 | 650327.517 | 244636.16 |

***Table A.10 – Descriptive statistics for the most relevant per workers variables (USD), post-Covid.***

| Variables | Group 1 | Group 2 | Group 3 | Group 4 | Group 5 | Group 6 |
|---|---|---|---|---|---|---|
| **property_plant_equip._net** | 508.487 | 464.2823 | 205.496 | 34.142 | 591.404 | 108.20 |
| **depreciation_amortization** | -73.935 | -124.0031 | -14.680 | -6.942 | -66.760 | -13.861 |
| **goods_sold_cost** | -306.516 | -262.8598 | -132.404 | -186.437 | -320.490 | -193.452 |
| **profit_net** | 108.063 | -79.5924 | 122.771 | 15.811 | 15.268 | 38.159 |
| **rd_expenses** | -37.761 | -12.9977 | -0.394 | -1.914 | -20.295 | -7.534 |
| **intangibles** | 258.338 | 255.8772 | 57.817 | 17.770 | 176.090 | 43.45 |

Descriptive statistics for other variables are available upon request for both the pre- and post-Covid period.

***Table A.11 - Changes in the cluster membership between pre- and post-Covid period.***

| ACF | | Post-Covid | | | | | |
|---|---|---|---|---|---|---|---|
| | | **Group 1** | **Group 2** | **Group 3** | **Group 4** | **Group 5** | **Group 6** |
| **Pre-Covid** | **Group 1** | - | 24 | 2 | 3 | 839 | 4 |
| | **Group 2** | - | 1 | 0 | 0 | 41 | 0 |
| | **Group 3** | 9 | - | - | 0 | 8 | 0 |
| | **Group 4** | 16 | 82 | - | - | 2522 | 14 |
| | **Group 5** | 756 | 38 | 4 | 3 | - | - |

***Table A.12 – Lasso regression, significant coefficients (full sample)***

| **Variables** | ***pre-Covid period*** | ***post-Covid period*** |
|---|---|---|
| inventory | 13.6241 | 5.0104 |
| assets_other | -1.1347 | . |
| property_plant_equipment_net | -0.5146 | -0.3269 |
| intangibles_other | -0.0223 | . |
| liabilities_current_other | -1.1864 | . |
| liabilities_non_current | -0.2847 | . |
| debt_interest_long_term | -0.2243 | . |
| liabilities_non_current_other | -0.7077 | . |
| sales_gross | 3.0183 | 0.0628 |
| share_reserves | -0.2813 | . |
| goods_sold_cost | . | -0.9998 |

*Figure A.1 – Densities and boxplots, pre- and post-Covid periods.*

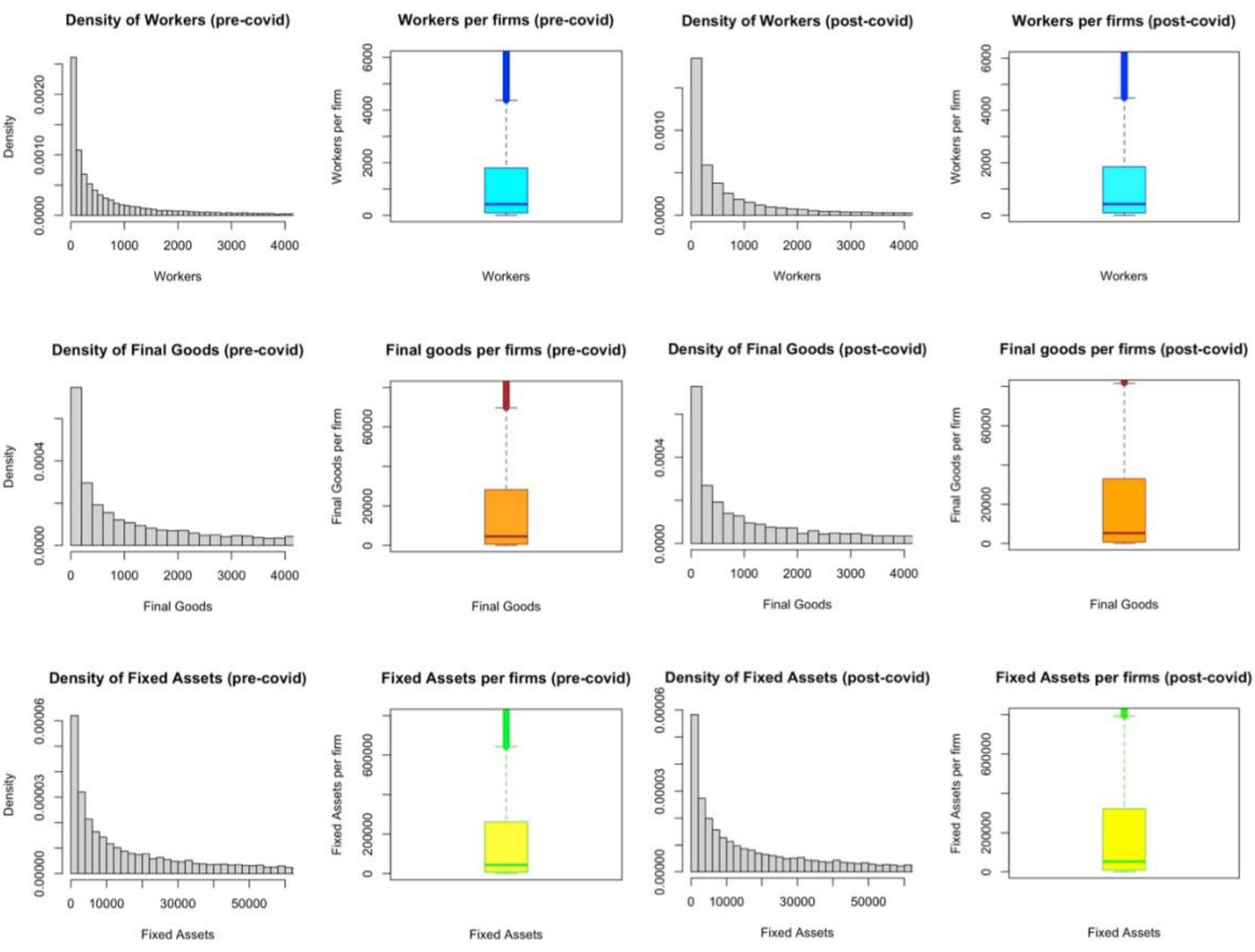

*Figure A.2 – Nodes distribution, pre- and post-Covid periods.*

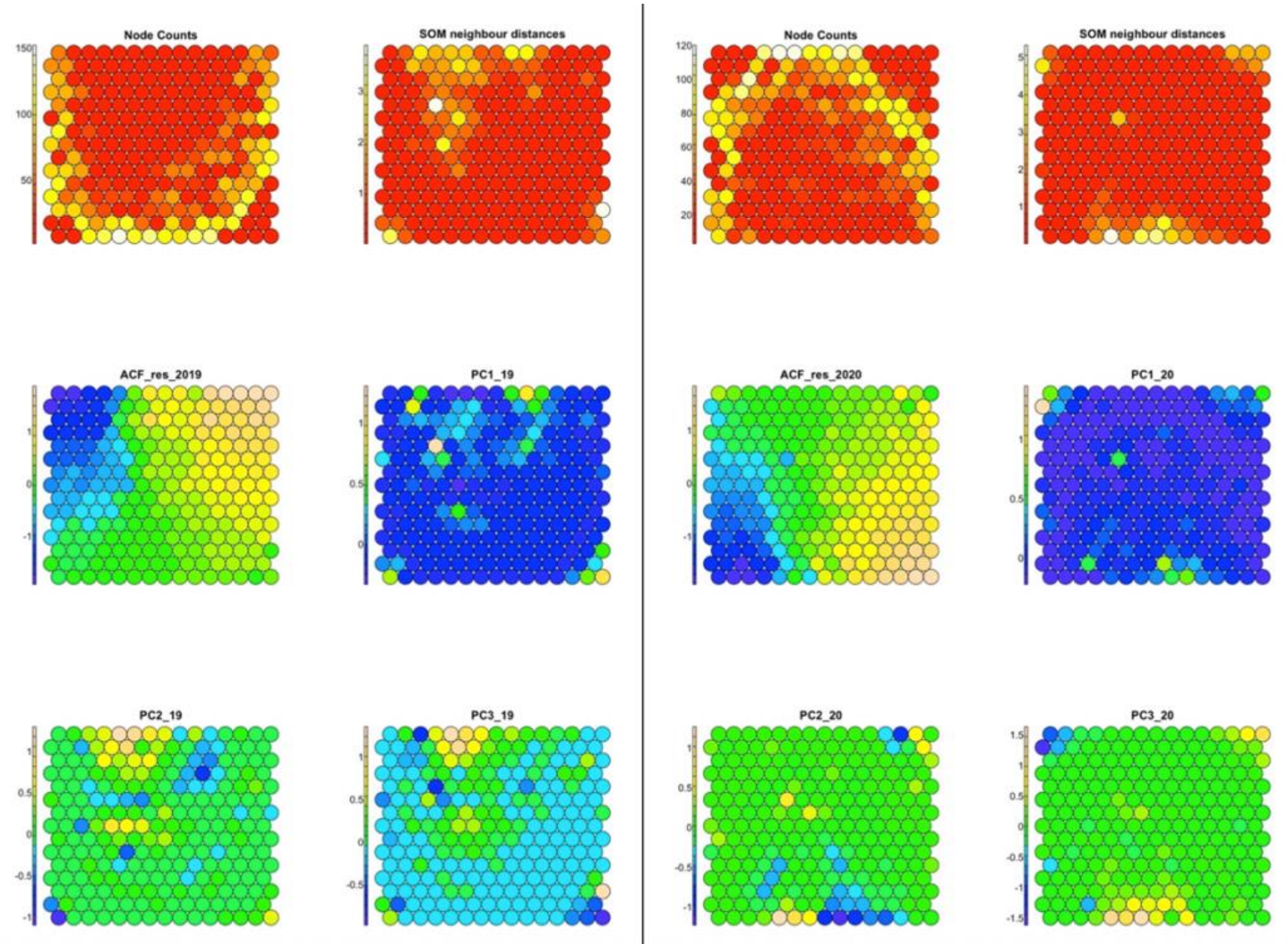